
\magnification=\magstep 1
\def\dot #1{{\mathaccent 95#1}}
\def\E{{\bf E}}
\def \D{{\cal D}}
\def\L{{\bf \Lambda}}

\def\x{{\bf x}}
\def\y{{\bf y}}
\def\A{{\bf A}}
\def\B{{\bf B}}
\def\D{{\cal D}}
\def\dd#1#2{{\delta\over{\delta {\bar A}\sp #1 ({\bar #2})}}}
\def\d#1#2{{\delta\over{\delta {\bf A }\sp #1 ({\bf #2})}}}
\def\ddd{{\delta\over{\delta{\bf A}}}}
\null\vskip 0.75in
{\hfil DTP-93-43}
\bigskip
\centerline{\bf Continuum Strong Coupling Expansion of Yang-Mills Theory:}
\centerline{\bf Quark Confinement and Infra-Red Slavery}

\vskip 0.5 in
\centerline{Paul Mansfield}
\bigskip
\centerline{Department of Mathematical Sciences,}
\centerline{University of Durham,}
\centerline{South Road,}
\centerline{Durham, DH1 3LE}

\vskip 0.5 in
\centerline{\bf Abstract}
\bigskip
\noindent
We solve Schr\"odinger's equation for the ground-state of
{\it four}-dimensional Yang-Mills theory as an expansion in inverse powers of
the coupling. Expectation values computed with the leading order
approximation are reduced to a calculation in {\it two}-dimensional Yang-Mills
theory which is known to confine. Consequently the Wilson loop in the
four-dimensional theory obeys an area law to leading order
and the coupling becomes infinite as the mass-scale goes to zero.

\vfill
\eject

\centerline{\bf Continuum Strong Coupling Expansion of Yang-Mills Theory:}
\centerline{\bf Quark Confinement and Infra-Red Slavery}
\bigskip
\centerline{\bf 1. Introduction}
\bigskip
Despite considerable success in describing high energy phenomena it
remains to be shown that Quantum Chromodynamics accounts for the known
and conjectured properties of the Strong Interactions at low energies such
as quark confinement and chiral symmetry breakdown. The strong coupling
expansion on the lattice [1], which leads to an
area law for Wilson loops, provides a framework
for the understanding of confinement but only in a calculational regime
where the continuous nature of space-time has been spoilt and the
connection with continuum field theory has only been established
numerically. The
difficulty with attempting a strong-coupling expansion
directly in the continuum is that our conventional
understanding of quantum field theory is based on
a different kind of expansion, namely
the semi-classical, or loop, expansion.
Indeed, the particle interpretation of field theory derives from
analysing the free part of the action in terms of harmonic oscillators.
This is encoded into
the LSZ formalism which underpins the use of field theory in Particle
Physics. In Yang-Mills theory asymptotic freedom [2] implies that the LSZ
realisation of field theory is applicable at high energies. As the energy
scale is reduced the running coupling grows but it is not known if it
continues to grow beyond the perturbative regime, although it is usually
thought to do so, giving rise to infra-red slavery.
In ordinary Quantum Mechanics the Born approximation, in contrast to the
semi-classical approximation, provides an expansion in decreasing powers
of $\hbar\sp 2$. In this paper we develop such a scheme for continuum
Yang-Mills theory in four dimensions. We will solve Schr\"odinger's
equation for the wave-functional of the ground-state as an expansion in
inverse powers of the coupling. Our strategy will be based on an {\it
ansatz} in which we assume that the logarithm of the wave-functional is
the integral of a local function of the gauge-potential. This enables us
to relate ground-state expectation values in the four-dimensional theory
to  calculations in two-dimensional Yang-Mills theory. It is well known
that Yang-Mills theory in two dimensions confines, and as a consequence,
our leading order approximation for the Wilson loop in the
four-dimensional theory has area-law behaviour.
Furthermore the coupling depends on an arbitrary mass-scale,
$\mu$, growing stronger as $\mu$ is reduced.
The relationship between
two-dimensional Yang-Mills theory and certain string theories has received
attention recently [3], so our result provides a link between this
and the four-dimensional theory.

\medskip
          The Yang-Mills action
can be written with the coupling factored out

$$ S_{YM}=-{1\over 2g\sp 2}\int d\sp 3 x\, dt\,tr\left(\E\sp 2-\B\sp
2\right),$$
$$\E=-{\dot \A}+\nabla A_0+[\A,A_0],\quad \B=\nabla\wedge\A+\A\wedge\A.
\eqno (1.1)$$
(The
gauge potentials are elements of the Lie algebra of the
gauge group so $\A=\A\sp CT\sp C,$ and $ (T\sp C)\sp \dagger=-T\sp
C,\,tr\,(T\sp AT\sp B)=-
\delta\sp {AB},\,[T\sp A,T\sp B]=f\sp {ABC}T\sp C$.) $g\sp 2$
plays the role of Planck's constant
since the action occurs in the path-integral as $exp\,iS/\hbar$ . Thus the LSZ
realisation of
Yang-Mills theory is appropriate to the semi-classical regime where
physical quantities are expandable as power series in $g\sp 2$. If we want to
understand the strong coupling limit of the theory we must expand in
powers of $1/g\sp 2$, in other words we need to study the ultra-quantum
mechanical limit where the (essentially semi-classical) notion of
particles may not be relevant. In this paper we will solve Schr\"odinger's
equation using such an expansion for the ground-state. Of course,
either kind of expansion should be viewed as being of a formal nature
since renormalisation leads to the introduction of a length scale in terms
of which the running coupling is fixed as a function of a mass scale.
Rather, the two formal power series should be viewed as different ways
of organising the calculation.

\medskip
    To simplify things let's consider Schr\"odinger's equation for a quantum
mechanical particle of mass $m$ moving in a potential $V(x)$

$$\left(-{\hbar\sp 2\over 2m}{d\sp 2\over dx\sp 2}+V(x)\right)\Psi={\cal E}
\Psi.\eqno (1.2)$$
If
we write the wave-function as $\Psi=exp\,-W(x)$ then this becomes

$$-{\hbar\sp 2\over 2m}\left( (W\sp \prime)\sp 2-W\sp {\prime
\prime}\right)+V={\cal E}.
\eqno (1.3)$$
In
the semi-classical approximation,
which corresponds to the loop expansion in the usual perturbative approach
to field theory, we expand

$$W=\sum\sp \infty_{n=0}\hbar\sp {n-1} W_n, \eqno (1.4)$$
so
that to leading order we can neglect $W\sp {\prime \prime}$ and obtain
the Hamilton-Jacobi equation

$$-{1\over 2m}(W_0\sp \prime)\sp 2+V={\cal E}, \eqno (1.5)$$
which
identifies $W_0$ as the classical action.
When the potential is sufficently weak we can use instead the Born
approximation.
Assuming that $V\rightarrow 0$ as $|x|\rightarrow \infty$ and that there
are no bound states implies that the lowest value of ${\cal E}$ is zero
corresponding to a (non-square-integrable) wave-function that is constant
at spatial infinity. If we take $\Psi =1-\psi$, with $\psi$ a small
perturbation, then re-arranging (1.2) gives

$$\psi=\left( -{1\over 2m}{d\sp 2\over dx\sp 2}+{V\over\hbar\sp 2}\right)\sp
{-1}
{V\over\hbar\sp 2}\eqno (1.6)$$
which can be expanded in powers of $V/\hbar\sp 2$.

\medskip
Another way of organising
the same expansion is to write $W$ as a power
series in $1/\hbar\sp 2$, $W=\sum\sp \infty_{n=1} \hbar\sp {-2n}{\tilde W_n}$
and expand (1.3). To
leading order and next to leading order this gives (with ${\cal E}=0$)

$${1\over 2m}{\tilde W}\sp {\prime \prime}_1+V=0,\quad \tilde W_2\sp
{\prime\prime}
=(\tilde W_1\sp \prime )\sp 2
.\eqno (1.7)$$
It is easy to check that this gives the same results to this order as
expanding (1.6).
This is the
approach we will take to the Schr\"odinger
equation for Yang-Mills theory.
\medskip
        We will follow the formulation of the Schr\"odinger equation for
Yang-Mills given by Jackiw in [4], which has the merits of being
both simple and closely
related to ordinary quantum mechanics.
(This reference also explains the modification to
the wave-functional required when there is a non-zero $\theta$-angle.)
We work in the Weyl gauge ($A_0=0$) so that the canonical
coordinates are $\A\sp C({\bf x})$ and their conjugate momenta are
$-g\sp {-2}\E\sp C({\bf x})$. The Hamiltonian is

$$H[\A,\E]=  -{1\over 2g\sp 2} \int d\sp 3x\,tr\left( \E\sp 2+\B\sp 2\right).
\eqno
(1.8)$$
Since
$\dot  A_0({\bf x})$ does not appear in the action the Euler-Lagrange
equation obtained by varying $A_0({\bf x})$,

$$\nabla \cdot\E+\A\cdot\E-\E\cdot\A=0,\eqno (1.9)$$
is a constraint rather
than a Hamiltonian equation of motion. In the
Schr\"odinger representation with $\hbar=1$
$\E\sp C$ is represented by the operator

$$\E\sp C ({\bf x})=ig\sp 2 \d C x , \eqno (1.10)$$
(showing
explicitly that $g\sp 2$ plays the role of Planck's constant).
Schr\"odinger's equation is

$$ \left(-{\textstyle 1\over 2}g\sp 2\Delta
+g\sp {-2}{\cal B}\right)\Psi[\A]=
{\cal E}\Psi[\A],$$
$$\Delta\equiv\int d\sp 3x \d C x\cdot\d C x,\quad
{\cal B}=-{\textstyle 1\over 2}\int d\sp 3x\,tr\,\B\sp 2
 ,\eqno (1.11)$$
and the
constraint is imposed as

$$\Gamma \,\Psi[\A ]=0 ,\quad \Gamma\equiv\nabla\cdot \d C x+f\sp {RSC}\A\sp R
\cdot\d S
x  .\eqno (1.12)$$
Now
$\Gamma$ is the generator of infinitesimal gauge transformations acting on
functionals of $\A$ so this constraint simply implies that the
wave-functional is gauge invariant. The Schr\"odinger equation stands in
need of regularisation because the two functional derivatives act at the
same point in space. In the next section we shall construct a regulated
operator, $\Delta_\phi$, that
preserves the gauge invariance and underlying geometry of the Hamiltonian
operator. It acts on integrals of local functions of $\A$ to produce other
local integrals with finite coefficents, unlike $\Delta$ which produces
coefficents that are ill-defined. This means that we have a Hamiltonian
operator that has a finite action on the space of sums of products of
integrals of local functions of $\A$ with finite coefficents. The price
for being able to do this is, as always, the introduction of an
arbitrariness in the renormalisation procedure that, in this case, is
encoded into a function, $\phi$. This is needed to separate
the finite part from quantities that diverge as we remove the cut-off.
If the theory is renormalisable, as we hope, we will be able to absorb this
arbitrariness into the coupling constant so that physical quantities
will not depend on $\phi$. Thus the finite version of the Schr\"odinger
equation will be

$$\left(-{1\over2}  g\sp 2\Delta_\phi
+{1\over g\sp 2}{\cal B}\right)\Psi[\A]=
{\cal E}\Psi[\A]\eqno (1.13)$$
where
$g$ is now a functional of $\phi$. If
we attempt a strong-coupling expansion of the ground-state
wave-functional

$$\Psi [\A ]\equiv exp -S[\A]=exp\,\left(-\sum_{n=1}\sp \infty
g\sp {-4n}S_n[\A ]\right),\quad {\cal E}=\sum_{n=1}\sp \infty g\sp {2-4n}{\cal
E}_n
\eqno (1.14)$$
the Schr\"odinger equation becomes

$${\textstyle 1\over 2}g\sp 2\Delta_\phi S-{\textstyle 1\over 2}g\sp 2\left(
\int d\sp 3x\,{\delta S\over\delta\A\sp C}\cdot{\delta S\over\delta\A\sp
C}\right)
+{1\over g\sp 2}{\cal B}
={\cal E}\eqno (1.15)$$
so that
to lowest order

$$\Delta_\phi S_1=2\left({\cal E}_1-{\cal B}\right). \eqno (1.16)$$
The right-hand side is the integral of a gauge invariant function that
is local in the sense of being constructed from fields at the point
${\bf x}$ and a finite number of derivatives. By construction, $\Delta_\phi$
will act on
integrals of local gauge-invariant functions to produce other integrals of
local gauge-invariant fuctions with finite coefficents. So we can look for
a solution for $S_1$ that is itself the integral of a local function. In
the next section we will show that

$$\Delta_\phi {\cal B}=\beta_1\left(\int d\sp 3x \right)
+\beta_2{\cal B}, \eqno (1.17)$$
where,
for the gauge group $SU(N)$ the coefficents are
$\beta_1=\phi\sp {(5)}(0)\,(N\sp 2-1)/320{\sqrt{\pi}}\sp 3$, and
$\beta_2=-11\phi\sp \prime (0)\,N/12{\sqrt{\pi}}\sp 3$. So if we take

$$S_1=-{2{\cal B}\over\beta_2},\quad
{\cal E}_1=-{\beta_1\over \beta_2}\int d\sp 3x, \eqno (1.18)$$
then
equation (1.16) is satisfied.
The infra-red divergent integral over space in ${\cal E}_1$ is what we
would expect for the energy of the ground-state, since by translational
invariance the energy density must be constant.
More generally we adopt an {\it ansatz}
of locality as a
strategy for solving the Schr\"odinger equation in a strong coupling
expansion.  We will assume that $S[\A]$  is a sum of integrals of local
functions of $\A$.
Then, by construction $\Delta_\phi S$ will also be a local integral, and so
too will the other terms entering Schr\"odinger's equation
in the form (1.15).
Taking $S$ to be gauge-invariant
satisfies the constraint (1.12), and taking the integrals to be invariant
under spatial translations and rotations acting on $\A$ ensures
that we are studying
the ground-state which can carry neither momentum
nor angular momentum.
We also require that the ground-state wave-functional be Lorentz invariant.
This gives a
very different wave-functional from that given by the semi-classical expansion
which would contain propagators, and so be non-local, but then we should
not expect the strong-coupling limit to resemble the semi-classical
one. To leading order the
ground-state wave-functional is

$$\Psi[\A]=exp\,{1\over 4\gamma\sp 2}\int d\sp 3x \,tr \B\sp 2,
\quad \gamma\sp 2={11 g\sp 4\phi\sp \prime (0)N\over 48\sqrt{\pi}\sp 3} \equiv
g\sp 4\kappa .\eqno (1.19)$$

\medskip
    Now that we have an approximation for $\Psi$ we can compute
expectation values of operators $\Omega[\A,\E]$.
We will be principally interested in the Wilson loop, in which case
$\Omega$ can be taken independent of $\E$, so $\Omega=\omega [\A]$,
say.(More generally
we use the Schr\"odinger
representation for $\E$, (1.10) to reduce the action of $\Omega$
on $\Psi$ to multiplication by some functional dependent on $\Psi$.)

$$\langle \Omega \rangle={\int {\cal D}\A\,\Psi\sp *[\A]\,\omega[\A]\,
\Psi[\A]\over
\int{\cal D}\A\,\Psi\sp *[\A]\,\Psi[\A]},\eqno (1.20)$$
and
to leading order this is

$$\langle\Omega\rangle\simeq{\int{\cal D}\A\,\omega[\A]\,exp
\left({1\over 2\gamma\sp 2}\int d\sp 3x
\,tr \B\sp 2\right)\over\int{\cal D}\A\,exp\left(
{1\over 2\gamma\sp 2}\int d\sp 3x \,tr\B\sp 2\right)}, \eqno (1.21)$$
But this is the functional integral for the vacuum expectation
value of $\omega$ in three dimensional Yang-Mills theory, Wick
rotated to Euclidean space,
and with coupling $\gamma$. We are back to the problem we started with
but in one dimension lower.
Feynman has discussed the qualitative structure of the wave-functional for
three-dimensional Yang-Mills theory, arguing that there is a mass gap [5].
We want to calculate (1.21) for large $g$,
this means that we can repeat the above
analysis. We begin by undoing the Wick rotation and working
in three-dimensional Minkowski space with time coordinate $x\sp 3$, and
quantisation surface $x\sp 3=constant$. We will
denote vectors in the 12-plane
by a bar. Working in the Weyl gauge $A_3=0$
the wave-functional of the vacuum is a functional of the two components
of the gauge potential $A_1, A_2$, and in the Schr\"odinger representation

$$\nabla_3 \bar A\sp C (\bar x)=-i\kappa g\sp 4\dd C x\eqno (1.22)$$
The
Schr\"odinger equation is

$$-{\textstyle 1\over 2}\int d\sp 2x\left(\kappa g\sp 4\dd C x \cdot \dd C x +
{1\over\kappa g\sp 4}tr \,B\sp 2
\right)\Phi[\bar A]= {\tilde{\cal E}}\Phi[\bar A],$$
$$ B=\epsilon \sp {ij}\left(\nabla_iA_j+A_iA_j\right)\eqno (1.23)$$
as
well as a constraint that demands that $\Phi$ be gauge-invariant.

$$\tilde\Gamma \,\Phi[\bar A ]=0 ,\quad \tilde\Gamma
\equiv\bar\nabla\cdot \dd C x+f\sp {RSC}\bar A\sp R\cdot\dd S x \eqno (1.24)$$
We
again construct a Hamiltonian operator that has a finite action on a space
of sums of products of local integrals with finite coefficents.
Consequently the
regulated Laplacian depends on a new arbitrary function, $\tilde\phi$, and
with $\tilde{\cal B}=-{\textstyle 1\over 2}\int d\sp 2x\,tr\,B\sp 2$,
Schr\"odinger's equation becomes

$$\left ( -{\textstyle 1\over 2} \gamma\sp 2\tilde \Delta_{\tilde\phi}
+{1\over \gamma\sp 2}{\tilde{\cal B}}\right)\Phi[\bar A]=
\tilde{\cal E}\Phi[\bar A]. \eqno (1.25)$$
Expanding
in a power series in $g\sp {-8}$
$$\Phi [\bar A ]\equiv exp -\tilde S[\bar A]=exp\,\left(-\sum_{n=1}\sp \infty
g\sp {-8n}\tilde S_n[\bar A ]\right),\quad \tilde{\cal E}
=\sum_{n=1}\sp \infty g\sp {4-8n}\tilde{\cal E}_n
\eqno (1.26)$$
we
obtain to leading order
$$\kappa\tilde \Delta_{\tilde\phi}\tilde S_1=
2(\tilde{\cal E}_1-\tilde{\cal B}/\kappa). \eqno (1.27)$$
Again
$\tilde \Delta_{\tilde\phi}$ acts on integrals of local gauge-invariant
functions to generate further integrals of local gauge-invariant functions,
and in the next section we compute

$$\tilde \Delta_{\tilde\phi}\tilde{\cal B}=
\tilde\beta_1\left(\int d\sp 2x\right)+\tilde\beta_2\tilde{\cal B}
\eqno (1.28)$$
With $\tilde\beta_1=(N\sp 2-1)\tilde\phi\sp {(2)}(0)/8\pi,\,\tilde\beta_2
=-23N/6\pi$ for $SU(N)$.
Taking
$$\tilde S_1=-{2\over \tilde\beta_2\kappa\sp 2}\tilde{\cal B},\quad
\tilde{\cal E}_1=-{\tilde\beta_1\over\tilde\beta_2\kappa}\left(\int d\sp
2x\right)
\eqno (1.29)$$
yields a
solution to (1.27).
The leading order expression for $\Phi[\bar A]$ is thus
$$\Phi[\bar A]=exp{1\over 4\gamma\sp 2_*}\int d\sp 2x\, tr\, B\sp 2,\quad
\gamma_*\sp 2={23\gamma\sp 4N\over 24\pi}
\eqno (1.30)$$
If
we restrict our attention to those $\omega[\A]$  that can be expressed as
functionals of only $\bar A$ on the quantisation surface
i.e. $\omega[\A]=\tilde\omega [\bar A ]$ then

$$\langle\Omega \rangle=
{\langle\Psi|\,\Omega \,|\Psi\rangle\over
\langle\Psi |\Psi\rangle}\simeq
{\langle\Phi|\,\tilde\omega \,|\Phi\rangle\over
\langle\Phi |\Phi\rangle}\eqno (1.31)$$
so
to leading order in the strong-coupling expansion

$$\langle \Omega \rangle\simeq{\int {\cal D}\bar A\,\Phi\sp *[\bar A]\,
\tilde\omega[\bar A]\,\Phi[\bar A]\over
\int{\cal D}\bar A\,\Phi\sp *[\bar A]\,\Phi[\bar A]}
\simeq {\int{\cal D}\bar A\,\tilde\omega[\bar A]\,exp\left(
{1\over 2\gamma\sp 2_*}\int d\sp 2x
\,tr B\sp 2\right)\over\int{\cal D}\bar A\,exp\left(
{1\over 2\gamma\sp 2_*}\int d\sp 2x \,tr B\sp 2\right)}, \eqno (1.32)$$
This is a functional integral for Yang-Mills theory in two
Euclidean dimensions so we are back to where we
started, but in two dimensions lower. We will
now stop this process of reducing the dimension,
because the gauge-invariant two-dimensional theory
is free in the gauge $A_2=0$ and we can evaluate (1.32) directly.

\medskip
  It is well known that Yang-Mills theory confines in two dimensions
in the sense that the Wilson loop
has an area law behaviour. The Wilson loop is the vacuum expectation value
of the trace of the path-ordered exponential of the gauge potential
$A$ taken around a closed curve [1]. If the curve is a rectangle of sides $R$
and $T$ then
in the Euclidean formulation of the theory this
has the interpretation of being
$exp-E(R)T$, where $E$ is the energy of a pair of massive `test-quarks'
created for a time $T$ and held a distance $R$ apart. With an area-law
behaviour $E(R)T=\sigma RT$
so the potential energy
of the two static quarks is $\sigma R$, requiring an
infinite energy to completely separate them.
We will now show how the Wilson loop for four dimensional Yang-Mills
theory shows area-law behaviour in the strong-coupling limit. In the Euclidean
formulation of the theory we want to compute

$$ W[C]={\int {\cal D}A_\mu \,
\left(tr\,P\,exp\left(-\oint_C dx\cdot A\right)\right)
exp\left({1\over 4g\sp 2}\int d\sp 4 x\, tr\, F_{\mu \nu} F_{\mu\nu}\right)
\over{\int {\cal D}A_\mu \,
exp\left({1\over 4g\sp 2}\int d\sp 4 x\, tr\, F_{\mu \nu}
F_{\mu\nu}\right)}}\eqno (1.33)$$
$P$ denotes path-ordering of the Lie algebra
generators around the closed curve $C$.
If $C$ is planar we can use
rotational invariance in four-space to place
it in the 12-plane. Then $W[C]$
is given by (1.20) with $\omega=tr \,P\,exp-\oint_Cd{\bf x}
\cdot\A$ and to leading order by (1.32)
with $\tilde\omega=tr\,P\,exp-\oint_Cd{\bar x}\cdot\bar A$.
Evaluating this in the gauge $A_2=0$ gives

$$W[C]\simeq
{\int{\cal D}A_1\,
\left(tr\,P\,exp\left(-\oint_C dx\sp 1 A_1\right)\right)
\,exp\left({1\over 2\gamma\sp 2_*}\int d\sp 2x
\,tr\, (\partial_2 A_1)\sp 2\right)\over\int{\cal D} A_1
\,exp\left({1\over 2\gamma\sp 2_*}\int d\sp 2x \,tr\, (\partial_2 A_1)\sp
2\right)}$$

$$=tr\,P\,exp\left(-{\gamma_*\sp 2\over 2}\oint_C\oint_Cdx\sp 1dy\sp 1
{\cal G}({\bar x},{\bar y})T\sp AT\sp A\right),\eqno (1.34) $$
where
${\cal G}$ is the Green's function for $\partial_2\sp 2$, i.e.

$$\partial_2\sp 2{\cal G}(\bar x ,\bar y )=\delta (\bar x ,\bar y ).
\eqno (1.35)$$
Bose
symmetry requires that $\cal G$ be symmetric in its arguments, so
we take

$${\cal G}(\bar x ,\bar y )={\textstyle 1\over 2}|x\sp 2-y\sp 2|\,
\delta (x\sp 1-y\sp 1).\eqno (1.36)$$

For
the Lie algebra $su(N)$ we have $T\sp AT\sp A=-1(N\sp 2-1)/N$. Because
this is proportional to the identity and because of the
$\delta (x\sp 1-y\sp 1)$ in the Green's function, the path-ordering is
redundant.
Furthermore the area enclosed by ${\cal C}$ can be written as

$${\cal A}[C]=-\oint \oint dx\sp 1\,dy\sp 1 {\textstyle 1\over 2}|x\sp 2-y\sp
2|\,\delta
(x\sp 1-y\sp 1). \eqno (1.37)$$
Putting
all this together gives the leading order approximation for the
Wilson loop as

$$W[C]\simeq
N\,exp\,\left(-\sigma {\cal A}[C]\right),\quad \sigma={(N\sp 2-1)
\gamma\sp 2_*\over 2N}=
\lambda N\sp 2(N\sp 2-1)(\phi\sp \prime (0))\sp 2g\sp 8.
\eqno (1.38)$$
Where $\lambda$ is a numerical constant, approximately equal to $2.6\times
10\sp {-4}$.
Now $\sigma$ is a physical quantity which must not depend on the arbitrary
function $\phi$, so the coupling must actually be a functional of $\phi$.
$\phi$ is dimensionless, but its argument has the dimension
of [length], so $\phi\sp \prime (0)\equiv \mu$ has the dimension of [mass].
Since our expansion is in increasing powers of $1/g\sp 4$ we shall define
a $\beta$-function

$$\beta\equiv\mu{\partial\over\partial\mu}\left({1\over g\sp 4}\right)
\eqno (1.39)$$
where
the differentiation is taken by holding physical quantities such as
$\sigma$ fixed. Applying this to the above expression for $\sigma$
gives

$$\beta={1\over g\sp 4},\eqno (1.40)$$
which
tells us that $1/g\sp 4(\mu)$ decreases, as the mass scale is reduced
(which we can also see directly from (1.38)).
We should emphasise at this point that our expansion in powers of
$1/g\sp 4$ is of the ground-state wave-functional, and not the physical
quantity $\sigma$. Thus if we were to compute the first correction
to (1.38) and use this to calculate the Wilson loop it would give corrections
to
$\sigma $ of higher order in $1/g\sp 4$, as well as, possibly, a contribution
of the same order in $g$ as our leading approximation, (1.38).
It is inevitable in any quantum theory
with only a single dimensionless coupling, where dimensionful physical
quantities acquire those dimensions through dimensional transmutation
that those quantities cannot be obtained from perturbation series in
powers of the coupling, or powers of any function of the coupling,
$f(g)$ say.
All physical masses, for example, must depend on $f$
and the arbitrary mass scale in the same way, i.e.
as $\mu\,exp\,-\int\sp f df\sp \prime
/\beta_f (f\sp \prime )$, with
$\beta_f =\mu{\partial f\over\partial\mu}$,
in order to be independent of $\mu$.
This would mean that if we could expand two physical masses
in power series in $f$ we would only need to compute the first terms
in order to know the ratio of the two masses to all orders.
Put another way if $f\rightarrow 0$ as $\mu\rightarrow\mu_*$ say,
then the $\mu$-independent ratio could be computed in this limit from
only the first terms in the series.
Such a
situation could not take into account the full non-linearity of the theory.
That our `strong-coupling' expansion of the ground-state wave-functional
gives a non-zero result for $\sigma$ tells us that necessarily this
will not be the first term in a power series expansion in the coupling.
We will not be despondent about this, as the alternative is an approximation
scheme in which we cannot compute $\sigma$ at all. However this may make
the renormalisation of the expansion difficult.

\medskip

    The wave-functional (1.19) is Lorentz invariant. Under a  Lorentz boost
the gauge condition $A_0=0$ must be preserved, so the change in $\A$ under
an infinitesimal transformation has two pieces, the first being the usual
one appropriate to the spatial components of a four-vector, whilst
the second is a gauge transformation that preserves the Weyl gauge. The
generator is thus $L({\bf\alpha})=i{\bf\alpha}\cdot\int d\sp 3x\,\x
{\cal H}$, with ${\cal H}$ the Hamiltonian density. This can be regulated
in the same way as $H$ in the next section. Now, by construction,
$H\Psi={\cal E}\Psi$, so ${\cal H}\Psi =(\nabla\cdot{\bf J}+const.)\Psi$,
for some ${\bf J}$. If we take $\int d\sp 3 x\,\x =0$, then $L({\bf\alpha})
\Psi=-(\int d\sp 3x\,{\bf\alpha}\cdot{\bf J})\Psi$. ${\bf J}$ has dimensions
of $[$length$]\sp {-3}$, and to the order we are working this can only come
from ${\cal H}$ acting on ${\cal B}/\mu$, so ${\bf J}$ must be $\mu\sp {-1}$
multiplied by a vector of dimension $[$length$]\sp {-4}$, but we cannot
construct a local gauge-invariant vector with this dimension out of $\A$
and positive powers of derivatives of $\phi$, so ${\bf J}$ must vanish,
as can be seen by explicit calculation, so $L({\bf\alpha})\Psi=0$.

\bigskip
\centerline{\bf 2. Heat-Kernel Regulator}
\bigskip
     We will now address the crucial question of how to
regulate Schr\"odinger's equation (1.11). The problem is
that it contains a product of two functional derivatives
at the same spatial point.
So, for example $\Delta$ acting on $-tr\,\A\sp 2(\x )$ is proportional to
$\delta (0)$
which is meaningless. More generally, if $\Sigma$ denotes the space of
sums of products of three-dimensional integrals of local functions of $\A$
with finite coefficents, then $\Delta$ acting on a typical element of
$\Sigma$ does not yield another element of $\Sigma$.
We will replace $\Delta$ in Schr\"odinger's equation
by a regulated operator, $\Delta_\phi $, which does have a finite action on
elements of $\Sigma$ and maps $\Sigma$ to itself. We will insist that for
those $\sigma\in\Sigma$ for which $\Delta\sigma\in\Sigma$ the action of
$\Delta$ and $\Delta_\phi$ coincide. In addition we require that
$\Delta_\phi$ is invariant under gauge transformations, as well as more
general transformations of coordinates on configuration space, consistent
with the interpretation of $\Delta$ as a Laplacian. This regularisation
and renormalisation procedure introduces an arbitrary function $\phi$ on
which $\Delta_\phi$ depends. This is an inevitable consequence of
isolating the finite part of a divergent quantity, and has the effect of
introducing an arbitrary scale into the problem. Ensuring that physical
quantities are independent of this arbitrariness, e.g. by absorbing it all
into the coupling, is the basic problem of renormalisation and has the
effect of making the coupling a functional of $\phi$. To construct
$\Delta_\phi$ we begin by `point-splitting' the functional derivatives.
Consider the operator

$$T_t\equiv\int d\sp 3x\,d\sp 3y\d R x \cdot
{\bf K}\sp {RS}({\bf x},{\bf y};t)\cdot
\d S y \eqno (2.1)$$
Where
the kernel, ${\bf K}$, is a second
rank tensor in three-space, satisfying some generalised heat equation
((2.16) below)
so that it depends smoothly on the proper time $t$, with the initial
condition

$${Limit\quad t\downarrow 0}\quad{\bf K}\sp {RS}({\bf x},{\bf y};t)
=\delta({\bf x}-{\bf y})\,\delta\sp {RS}{\bf 1} \eqno (2.2)$$
it
has the interpretation of being the `temperature' at the point $\x$ at time
$t$ due to a delta-function source placed at $\y $ at the initial time.
$\bf 1$ is the identity tensor (dyadic) in three-space.
Taking $t$ small but  non-zero, $t=\epsilon$ say, gives a regulated
operator that acts on elements of $\Sigma$ to produce other elements of
$\Sigma$. $T_\epsilon$ acting on $\sigma\in\Sigma$
produces local integrals because in the small time $\epsilon$ heat cannot flow
vary far from its original source and so ${\bf K}(\x ,\y ,\epsilon ) $
will depend only on $\A  (\x)$  for  $\x$ in a neighbourhood of $\y $
of volume of order ${\sqrt\epsilon}\sp 3$. However, the coefficents of the
integrals
resulting from the action of $T_\epsilon$ on $\sigma\in\Sigma$ will
contain powers of $\epsilon$ some of which diverge as $\epsilon\rightarrow
0$, so we cannot simply replace $\Delta\sigma$ by the $Limit\quad\epsilon
\rightarrow 0\quad T_\epsilon\sigma$. These powers of $\epsilon$ may be
determined from
dimensional analysis. We will see that the proper-time $\epsilon$ has
dimensions of $[$length$]\sp 2$, whilst $T$ has dimensions of $[$length$]\sp
{-1}$.
We will choose ${\bf K}$ to preserve gauge and rotational invariance,
so, given the local character of $T_\epsilon$, we have, for example

$$T_\epsilon{\cal B} =$$
$$\int d\sp 3x\left( \alpha_1
\epsilon\sp {-5/2}+\alpha_2\epsilon\sp {-1/2} tr\B\sp 2
+\alpha_3\epsilon\sp {1/2}tr\left((\nabla\wedge\B+\A\wedge\B+\B\wedge\A)\sp 2
\right)+... \right) ,\eqno (2.3)$$
where
the $\alpha_i$ are numerical constants.
We can extract a finite result for
$T_\epsilon\sigma$ when $\epsilon =0$ using analytic continuation
in an analogous fashion to zeta-function regularisation [6].
Suppose $\phi (p)$ is a differentiable function equal to one at the
origin, and that $f(p)$ is also a differentiable function which is finite
at the origin and such that $\phi (p) f(p)$ vanishes at infinity. Then on
integrating by parts

$$I(s)\equiv\int_0\sp \infty dp\,p\sp {s-1}\phi (p)\, f(p)=
\left [  {p\sp s\over s}\phi (p)\,f(p)\right ]_0\sp \infty -\int_0\sp \infty
dp \,{p\sp s\over s} {d\over dp}(\phi f).\eqno (2.4)$$
For
$s>0$ the first term vanishes, so

$$Limit\quad s\downarrow 0\quad\quad sI(s)=
-\int_0\sp \infty dp\, {d\over dp}(\phi f)=f(0).\eqno (2.5)$$
If
$f$ diverges at the origin like $p\sp {-n}$, with $n$ an integer, then $f(0)$
is meaningless, but we can give a meaning to the left-hand side of (2.5)
and use this as our definition of $f(0)$. The integral $I(s)$ exists
(provided $\phi f$ has no other divergences) for $s>n$, so if we
analytically continue $sI(s)$ to small values of $s$ we can take the limit
as $s\downarrow 0$ to obtain a finite result. Thus, repeated integration
by parts for $s>n$ gives

$$sI(s)=-\int_0\sp \infty{p\sp s\over (n-s)(n-s-1)..(1-s)}{d\sp {n+1}\over
dp\sp {n+1}}
\left(\phi (p)\, f(p)\, p\sp n\right).\eqno (2.6)$$
This
expression allows an analytic continuation to $s<n$ so we can take
the limit to obtain

$$Limit \quad s\downarrow 0 \quad\quad s I(s)={1\over n!}{d\sp {n}\over dp\sp
{n}}
\left(\phi f p\sp n\right)|_{p=0}. \eqno (2.7)$$
Clearly
this result depends on the arbitrary function $\phi$, as we should expect.
Extracting a finite quantity from a divergent one requires the use of an
arbitrary renormalisation procedure in general and the introduction of an
arbitrary scale in particular.  Physical quantities will be independent
of $\phi$ if the theory is renormalisable. With the use of this tool we
take the action of $\Delta_\phi$ on an arbitrary element, $\sigma\in
\Sigma$ to be

$$\Delta_\phi \sigma=\quad Limit\quad s\downarrow 0\quad s\int_0\sp \infty
dp\,p\sp {s-1}\phi (p) \, T_{p\sp 2}\sigma, \eqno (2.8)$$
where
the integral on the right-hand side is defined by analytic continuation in
$s$ as in the example above. We take $T$ at proper-time $p\sp 2$ rather than
$p$ because of the square-roots appearing in (2.3) which are generic to a
theory in an odd number of dimensions, (by which we mean here the three
spatial dimensions). For example

$$\Delta_\phi {\cal B}=\int d\sp 3x\left({\alpha_1\over 5!}\phi\sp {(5)}(0)
+\alpha_2\phi\sp \prime (0) \,tr\, \B \sp 2 \right). \eqno (2.9)$$
More
generally, $\Delta_\phi$ acting on a local integral of dimension
$[$length$]\sp {-D}$ results in other local integrals
of dimensions $[$length$]\sp {-D\sp \prime}$ with $D\sp \prime\le D$. We have
done
more than regulate $\Delta$ by constructing an operator whose action on
elements of $\Sigma$ does not depend on a cut-off. We have actually done
part of the job of renormalisation. The rest, ensuring that physical
quantities are independent of $\phi$, remains to be done.
\medskip
   We
now turn to the choice of heat-kernel. Under
an infinitesimal gauge transformation parametrised by $\omega$

$$\A \sp C({\bf x})\rightarrow\A \sp C ({\bf x})+\nabla\omega\sp C({\bf
x})-f\sp {RSC}
\omega\sp R ({\bf x})\A\sp S({\bf x}),\eqno (2.10)$$
the
functional derivative transforms homogeneously as

$$\dd C x \rightarrow \dd C x -f\sp {RSC}\omega\sp R( \x )\dd S x ,\eqno
(2.11)$$
so that
$\Delta$ is invariant. We
will demand that $\Delta_\phi$ is also gauge-invariant, which in turn
requires that  $T_t$ be gauge-invariant, so we insist that the
kernel transform under (2.10)
as

$${\bf K}\sp {RS}({\bf x},{\bf y};t)\rightarrow
{\bf K}\sp {RS}({\bf x},{\bf y};t)-
f\sp {ABR}\omega\sp A({\bf x}){\bf K}\sp {BS}({\bf x},{\bf y};t)
-{\bf K}\sp {RB}({\bf x},{\bf y};t)f\sp {ABS}
\omega\sp A({\bf y}).\eqno (2.12)$$
Or
in matrix notation with $\omega=(\omega\sp {RS})\equiv(\omega\sp Qf\sp {QRS})$,

$${\bf K}({\bf x},{\bf y};t)\rightarrow
{\bf K}({\bf x},{\bf y};t)+
\omega({\bf x})\,{\bf K}({\bf x},{\bf y};t)
-{\bf K}({\bf x},{\bf y};t)\,
\omega({\bf y}).\eqno (2.13)$$
It
is easy to construct a kernel that satisfies these
conditions. If ${\bf\Lambda}$ is a local operator
that acts on variations of the gauge-potential as

$$(\Lambda\circ\delta \A)\sp C(\x)=\int d\sp 3y\,{\L\sp {CD}}(\x ,\y )\cdot
\delta\A\sp D(\y ),\eqno (2.14)$$
then
we can view ${\bf\Lambda}$ as a matrix with indices $\x,\y,CD$ as
well as the three-space tensor indices, and $\circ$ as matrix
multiplication. If, in addition, $\L$ transforms under (2.10) as

$$\L (\x ,\y )\rightarrow \L (\x ,\y ) +\omega (\x)\, \L (\x, \y)
-\L ( \x , \y )\, \omega (\y ), \eqno (2.15)$$
then
we can take the kernel to be $exp\,-t\L\circ$ acting on the identity
matrix. For example, if $\D$ is the covariant derivative acting
in the adjoint representation , i.e. $\D_i \delta A_j=
\nabla_i\delta A_j +[A_i,\delta A_j]$ then we could (but we won't)
take $\L\circ\delta\A=-\D_i\D_i\delta\A$, in which case
$\L (\x, \y )=\D_i\D_i\delta (\x ,\y )$ satisfies (2.15).
We can then compute $\bf K$ by solving the heat equation

$$-\L\circ{\bf K}={\partial\over\partial t}{\bf K}, \eqno  (2.16)$$
subject
to the initial condition (2.2).

\medskip
     There are many choices of operator $\L$ that satisfy the
above conditions. We need a further criterion to pick the right
one. This will emerge from considering the geometry that
underlies Schr\"odinger's equation (1.11). If we think of $\A(\x )$
as being a set of coordinates on the space of configurations, ${\cal A}$,
then this space has a natural (flat) metric,

$${ g}(\delta \A,\delta\A )=-\int d\sp 3 x\, tr\,\delta \A (\x )\cdot\delta
\A (\x ), \eqno (2.17)$$
which
is invariant under (2.10). We can use this metric to define the
volume element $\D \A$ in (1.20). Furthermore the operation $\circ$
is just the scalar product constructed using this metric.
The operator $\Delta$ is then the
Laplacian, expressed in terms
of the coordinates $\A$,
acting on functions on $\cal A$ such as $\Psi$.
i.e. if $D/D\A$ is the Levi-Civita connection on $\cal A$
(indistinguishable from $\delta/\delta\A$ for the metric (2.11))
then

$$\Delta =g\sp {-1}\left({D\over D\A},{D\over D\A}\right)=
{D\over D\A}\circ{D\over D\A}, \eqno (2.18)$$
and

$$T={D\over D\A}\circ {\bf K}\circ{D\over D\A}. \eqno (2.19)$$
Physics
should be independent of how we choose coordinates on $\cal A$,
so Schr\"odinger's equation must transform covariantly under
a change of coordinates $\A( \x )$.
$\Psi$ transforms as a scalar, and $\Delta $ is the
scalar Laplacian so ${\cal B}\equiv-{\textstyle 1\over 2}\int d\sp 3
x\,tr\,\B\sp 2$
must also transform as a scalar.
Now we want the regulated operator $T$ to act on scalars to
produce scalars, and this will be the case if ${\bf K}$
is a second rank tensor under transformations of $\cal A$.
This in turn will be guaranteed if $\L$ is also a second rank tensor.
So we need to find a local operator in three-space enabling us to
construct the heat equation (2.10), which is also a tensor
under coordinate transformations of ${\cal A}$. We can
construct tensors by differentiating scalars using the Levi-Civita
connection. The only scalar we have
identified so far is ${\cal B}$, so

$${D\over D\A}{\cal B}={\delta\over\delta\A}{\cal B}
=\nabla\wedge\B+\A\wedge\B+\B\wedge\A
\equiv\D\wedge\B \eqno (2.20)$$
is
a covariant vector on ${\cal A}$, whilst a suitable candidate for $\L$ is

$$ \L={D\over D\A}\otimes{D\over D\A}\,{\cal B}.\eqno (2.21)$$
With
the flat metric (2.11) this gives

$$\L\circ\delta\A= -\D\cdot\D\delta\A +\D\D\cdot\delta\A
+2\B\wedge\delta\A+2\delta\A\wedge\B$$

$$\equiv -\D\cdot\D\delta\A +\D\D\cdot\delta\A
-2{\bf F}\cdot\delta\A \eqno (2.22)$$
This
is just the operator that appears in the semi-classical
expansion of three-dimensional Yang-Mills. ${\bf F}$ is the
field-strength tensor in the adjoint representation.

\medskip
    We need to compute the coefficents $\alpha_1,\alpha_2 $
that occur in the expansion (16). Now

$$T{\cal B}=  \ddd\left(\circ{\bf K}\circ\left(\ddd  {\cal B}\right)\right)=
\left (\ddd \circ {\bf K}\right )\circ
\ddd {\cal B}
+Tr\left({\bf K}\circ\ddd\otimes\left(\ddd {\cal B}\right)\right).\eqno
(2.23)$$
$Tr$ denotes a trace over all the matrix indices.
The
first term on the right-hand side is proportional
to $\D\wedge\B$ and so contributes to $\alpha_3$
but not $\alpha_1$ and $\alpha_2$. We are left with

$$Tr\left({\bf K}\circ\ddd\otimes\left(\ddd {\cal B}\right)\right)=
Tr\left({\bf K}\circ\L\right)
=-{\partial\over\partial t}Tr\left({\bf K}\right)
\eqno (2.24)$$
Evaluated at $t=\epsilon$.
So
we need to compute the diagonal elements of ${\bf K}$. We could attempt
to do this for a general $\A$, but since we only
need to extract the numerical coefficents
$\alpha_1,\, \alpha_2$ we can do the calculation for
a background satisfying the classical equation
of motion $\D\wedge\B=0$. We will show that
in this case

$$exp\,\left({-t\L\circ}\right)=exp\,\left({t(\D\cdot\D{\bf 1}+2{\bf
F})\cdot}\right)
-\int_0\sp td\tau\,\D\,exp\,\left({\tau\D\cdot\D}\right)\D\cdot
\eqno (2.25)$$
({\bf K} can be found for a general configuration
by perturbing about this result in powers
of $\D\wedge\B$.) For an arbitrary $\A$

$$[\,\D\cdot\D,\D\,]=-2{\bf F}\cdot\D +(\D\cdot
{\bf F}),\eqno (2.26)$$
so
that

$$\L\circ\D\omega=-(\D\cdot{\bf F})\,\omega,\quad
\D\cdot\L\circ\delta\A=(\D\cdot{\bf F})\cdot\delta\A .
\eqno (2.27)$$
{}From
now on we will assume that $\A$ satisfies
$\D\wedge\B=0$ which implies that $\D\cdot {\bf F}$ vanishes
so then
$\D\omega$ is a zero-mode of $\L\circ$
and $\D\cdot(\L\circ\delta\A)=0$. Let $\cal P$
denote the projector onto the kernel of $\D\cdot$
so that

$${\cal P}\circ\delta\A=\delta\A
-\D\left(\D\cdot\D\right)\sp {-1}\D\cdot\delta\A, \eqno (2.28)$$
Now $\L\circ{\cal P}=\L$ so

$$exp\,\left({-t\L\circ}\right){\cal P}\circ\delta \A=
exp\,\left({-t\L\circ}\right)\delta\A-\D\left(\D\cdot\D\right)\sp {-1}\D
\cdot\delta\A,
\eqno (2.29)$$
If
${\bf u}$ is in the kernel of $\D\cdot$
then

$$\L\circ{\bf u}=-\D\cdot\D{\bf u}-2{\bf F}\cdot{\bf u}
\eqno (2.30)$$
which
will also be in the kernel of $\D\cdot$
so we can write

$$exp\,\left({-t\L\circ}\right){\cal P}=exp\,\left({t(\D\cdot\D {\bf 1}+
2{\bf F})\cdot}\right){\cal P}\eqno (2.31)$$
{}From
(2.26)

$$\left(\D\cdot\D {\bf 1}+2{\bf F}\right )\cdot\D
=\D\,\D\cdot\D, \eqno (2.32)$$
so

$$exp\,\left({t(\D\cdot\D {\bf 1}+
2{\bf F})\cdot}\right)\D=\D exp\,\left( {t \D\cdot\D}\right), \eqno (2.33)$$
(as
can be seen by expanding both exponentials in power series, and
using (2.32) to move $\D$ to the left.)
This enables us to simplify (2.31) to

$$exp\,\left({-t\L\circ}\right){\cal P}\circ=exp\,\left({t(\D\cdot\D {\bf 1}+
2{\bf F})\cdot}\right)-\D exp\,\left({t\D\cdot\D}\right)
\left(\D\cdot\D\right)\sp {-1}\D\cdot
\eqno (2.34)$$
Finally
we can write (2.29) as

$$exp\,\left({-t\L\circ}\right)=exp\,\left({t(\D\cdot\D {\bf 1}+
2{\bf F})\cdot}\right)-\D \,exp\,\left( {t\D\cdot\D}\right)\left(\D\cdot\D
\right)\sp {-1}\D\cdot
+\D\left(\D\cdot\D\right)\sp {-1}\D\cdot\eqno (2.35)$$
which
is equivalent to (2.25). Using cyclicity of $Tr$ we obtain

$${\partial\over\partial t}Tr\,{\bf K}=
{\partial\over\partial t}\left(Tr\,exp\,\left({t(\D\cdot\D {\bf 1}+
2{\bf F})\cdot}\right)- Tr\,exp\,\left({t\D\cdot\D}\right)\right).\eqno
(2.36)$$
We
can make this more explicit by introducing the heat-kernels

$$H\sp {CD}(\x ,\y ,t)\equiv \,exp\,\left( {t\D\cdot\D}\right)\sp {CD}\delta
(\x -\y
),$$
$${\bf G}\sp {CD}(\x ,\y , t)\equiv
\,exp\,\left({t(\D\cdot\D {\bf 1}+
2{\bf F})\cdot}\right)\sp {CD}\delta (\x -\y)\,{\bf 1},\eqno (2.37)$$
which
satisfy the generalised heat equations

$${\partial\over\partial t}H=\D\cdot\D H,\quad
{\partial\over\partial t}{\bf G}=(\D\cdot\D {\bf 1} +2{\bf F})\cdot
{\bf G},\eqno (2.38)$$
subject to
the initial conditions

$$Limit\quad t\downarrow 0\quad H\sp {BC}(\x ,\y )= \delta (\x -\y
)\delta\sp {BC}, $$

$$Limit\quad t\downarrow 0 \quad {\bf G}\sp {BC} (\x ,\y)=
\delta (\x -\y )\delta\sp {BC}{\bf 1}. \eqno (2.39)$$
In terms
of these functions the right-hand side of (2.36) becomes

$${\partial\over\partial t}\left (\int d\sp 3x\, G\sp {CC}_{ii}
(\x ,\x ,t)-\int d\sp 3 x\,H\sp {CC}(\x ,\x , t)\right), \eqno (2.40)$$
so
we need to know the diagonal matrix elements of ${\bf G}$
and $H$ for short times. These may be computed using a standard
technique [7] that assumes the existence of expansions of the form

$$H\sp {BC}(\x ,\y , t)
={e\sp {-(\x -\y )\sp 2/4t}\over {\sqrt{4\pi t}}\sp 3}\sum_{n=0}\sp \infty
t\sp n{a}_n\sp {BC}(\x ,\y ) ,$$

$$ {\bf G}\sp {BC}(\x ,\y , t)
={e\sp {-(\x -\y )\sp 2/4t}\over {\sqrt{4\pi t}}\sp 3}\sum_{n=0}\sp \infty
t\sp n{\bf b}_n\sp {BC}(\x ,\y ) .\eqno (2.41)$$
The
prefactor solves the ordinary heat equation and collapses
to a delta-function at small times, consequently the initial conditions
are satisfied by taking

$$a_0\sp {BC}(\x ,\x )=\delta\sp {BC}, \quad {\bf b}_0\sp {BC}(\x ,\x )
=\delta\sp {BC}{\bf 1}. \eqno (2.42)$$
Substituting the
expansions into the heat-equations (2.38) and
equating terms of the same order in $t$ leads to recurrence relations

$$(\x -\y )\cdot\D {\bf b}_0(\x ,\y )=0$$

$$(\D\sp 2{\bf 1}+2{\bf F})\cdot {\bf b}_{n-1}-(\x -\y )\cdot\D {\bf b}_n
(\x ,\y )=n{\bf b }_n (\x ,\y ), \quad n>0$$

$$(\x -\y )\cdot\D a_0(\x ,\y )=0,$$

$$\D\sp 2 a_{n-1}-(\x -\y )\cdot\D a_n
(\x ,\y )=na_n (\x ,\y ),\quad n>0.\eqno (2.43)$$
Repeated
applications of $\D$ to these equations, followed by setting $\x =\y$
enables them to be solved for ${\bf b}_n (\x ,\x ), a_n (\x ,\x )$ and
their derivatives. For example

$$a_1(\x ,\x )=0,\quad a_2(\x ,\x )={1\over 12}F_{ij} F_{ij},$$

$${\bf b}_1(\x ,\x )=2{\bf F},\quad
{\bf b}_2 (\x ,\x )=2{\bf F}\cdot{\bf F}+{1\over 12}{\bf 1}\,
F_{ij}F_{ij}+{1\over 3}\D\cdot\D {\bf F} \eqno (2.44)$$
Using
these in (2.40) and (2.23) gives eventually

$$\alpha_1={3n_g\over{\sqrt {4\pi}}\sp 3},\quad \alpha_2={11C\over 6{\sqrt
{4\pi}}\sp 3}\eqno (2.45)$$
where
$n_g$ is the number of Lie algebra generators and $C$ is given by
$f\sp {RPQ}f\sp {SQP}=-\delta\sp {RS}C$. For $SU(N)$ we have
$n_g=N\sp 2-1$ and $C=2N$.
Note that if we had performed this calculation for an Abelian theory, such
as QED, then $\Lambda$ would not depend on $\A$ and consequently the
right-hand side of (2.9) would just be independent of $\A$. We would then
be unable to solve (1.16) with our locality {\it ansatz}.

\medskip
       To construct the Schr\"odinger equation for
three-dimensional Yang-Mills theory we need an operator that
acts on an element, $\tilde\sigma$, of a space $\tilde\Sigma$ of
two-dimensional integrals of local functions of $\bar A$ to produce other
elements of that space. Proceeding as before we
define an operator ${\tilde\Delta}_{\tilde\phi}$ so that its
action on $\tilde\sigma$ coincides with that of the unregulated
Laplacian when the latter is finite. Thus

$${\tilde\Delta}_{\tilde\phi}{\tilde\sigma}=
\quad Limit\quad s\downarrow 0\quad s\int_0\sp \infty
dp\,p\sp {s-1}\tilde\phi (p) \, \tilde T_{p}\tilde\sigma,$$

$$\tilde T_t=\int d\sp 2x\,d\sp 2 y\,\dd R x \cdot \bar K \sp {RS}
(\bar x,\bar y,t)\cdot\dd S y .\eqno (2.46)$$
To ensure
gauge and reparametrisation invariance the
kernel is chosen as $\bar K=exp\,-t\bar\Lambda\circ$
with $\bar\Lambda$ a tensor in the space
of gauge-potentials,

$$\bar\Lambda\sp {RS}(\bar x ,\bar y )
\equiv\dd R x \dd S y \tilde{\cal B}.\eqno (2.47)$$
As
before, $\bar K$ can be expressed in terms of simpler heat-kernels

$$\bar K (\bar x , \bar y )=\tilde G (\bar x ,\bar y)-\int_0\sp t
d\tau \tilde\D \tilde H (\bar x , \bar y )\tilde\D\cdot,\eqno (2.48)$$
with $\tilde\D f=\bar\nabla f+[\bar A ,f]$ and

$${\partial\over\partial t}\tilde H=\tilde\D\cdot\tilde\D \tilde H,\quad
{\partial\over\partial t}\bar{ G}=(\tilde\D\cdot\tilde\D \bar{ 1} +2{\bar F})
\cdot
\bar{ G},\eqno (2.49)$$
Here $F_{ij}\equiv \epsilon_{ij} B_{adj}$ is the field strength in the
adjoint representation.
$\bar G$ and $H$ are assumed to have small $t$ expansions
of the form

$$ {\bar G}\sp {BC}(\bar x ,\bar y , t)
={e\sp {-(\bar x -\bar y )\sp 2/4t}\over {{4\pi t}}}\sum_{n=0}\sp \infty
t\sp n{\bar b}_n\sp {BC}(\x ,\y ) ,$$

$$H\sp {BC}(\bar x ,\bar y , t)
={e\sp {-(\bar x -\bar y )\sp 2/4t}\over {{4\pi t}}}\sum_{n=0}\sp \infty
t\sp n{\tilde a}_n\sp {BC}(\x ,\y ) .\eqno (2.50)$$
The calculation
of the expansion coefficents is the same as in the
previous case, leading to

$$\tilde a_1(\bar x ,\bar x )=0,
\quad \tilde a_2(\bar x ,\bar x )={1\over 6}B_{adj}\sp 2,$$

$${\bar b}_1(\bar x ,\bar x )=2{\bar F},\quad
{\bar b}_2 (\bar x ,\bar x )=-{11\over 6}B_{adj}\sp 2\bar 1+
{1\over 3}\tilde\D\cdot\tilde\D {\bar F} \eqno (2.51)$$
Using
these expansions we find

$$\tilde T_\epsilon\tilde{\cal B}=
\int d\sp 2x\left(\tilde\alpha_1\,\epsilon\sp {-2}+\tilde\alpha_2\,tr\,B\sp
2+...
\right)\eqno (2.52)$$
with $\tilde\alpha_1={n_g\over 4\pi},\quad\tilde\alpha_2={23C\over24\pi}$,
from which follows (1.28).

\vfill
\eject

\centerline{\bf 3. Conclusions}
\bigskip
   We have constructed a regulated form of Schr\"odinger's equation for
four-dimensional continuum Yang-Mills theory in which the Hamiltonian operator
has a
finite action on integrals of local functions of $\A$. We solved this
for the logarithm of the ground-state wave-functional as a power series
in $1/g\sp 4$ on the assumption that the terms in this series are
themselves integrals of local functions of $\A$. Ground-state expectation
values
calculated using the leading order result reduce to calculations in
two-dimensional Yang-Mills theory. This is known to confine and be
directly related to certain string theories. Consequently the Wilson loop
in the four-dimensional theory satisfies an area law, and the coupling is
seen to grow as the mass-scale, $\mu$, is reduced, which is infra-red
slavery.
It is the intrinsic non-linearity of Yang-Mills theory that is responsible
for this. We would not have been able to solve Schr\"odinger's equation
for an Abelian gauge theory
using the {\it ansatz} of locality.
Our results are
preliminary, and many questions remain to be answered. We need to
know how to ensure that all physical quantities are independent of the
arbitrary function $\phi$ that was introduced to
regulate, and partially renormalise the theory. We need to compute
corrections, and to construct the excited states. We also need to know how
to incorporate fermions. We expect that other quantum theories
such as $CP\sp n$-models and String Theories may be treated in a similar
fashion.

\bigskip
\noindent
Finally, it is a pleasure to acknowledge conversations with Ed Corrigan on
the subjects of zeta-function regularisation and Yang-Mills theory,
and also with Ryu Sasaki.

\bigskip
\bigskip
\centerline{\bf 4. References}

$$\vbox{\halign{#\hfil&\hskip 0.1in#\hfil\cr [1]& K.G.Wilson Phys.Rev.D
10(1974)2445\cr
[2]& D.J.Gross, F.Wilczek Phys. Rev. Lett. 30(1973)1343\cr
   & H.D.Politzer Phys. Rev. Lett. 30 (1973) 1346\cr
[3]& D.J.Gross, W.Taylor PUPT-1376 LBL-33458 UCB-PTH-93/02\cr
[4]& R.Jackiw Rev. Mod. Phys. 52(1980)661\cr
[5]& R.P. Feynman Nucl. Phys. B188(1981)479             \cr
[6]& E.Corrigan, P.Goddard, H.Osborn, S.Templeton Nucl. Phys.
B159(1979)469\cr
   & J.S.Dowker, R.Critchley  Phys. Rev. D13(1976)3224, D16(1977)3390\cr
   & S.Hawking Comm. Math. Phys. 55(1977)133\cr
   & D.B.Ray, I.M.Singer Adv. Math. 7(1971)145\cr
[7]& B.S. deWitt Phys. Reports 19(1975)295\cr
   & H.P.McKean, I.M.Singer, J. Diff. Geom. 5(1971)233\cr
   & P.B.Gilkey, J. Diff. Geom. 10(1975)601, Proc. Symp. Pure Math. 27
(1975)265\cr}}$$

\vfill
\eject
\bye